\documentclass[18pt,preprint] {aastex}
\usepackage{graphicx,epsfig,natbib,color}
\usepackage{lscape}
\usepackage{graphicx}
\usepackage{epsfig}
\usepackage{natbib}
\usepackage[colorlinks, citecolor=blue]{hyperref}

\slugcomment{Accepted by SCIENCE CHINA Earth Sciences}

\begin{document}
\title{Origin and Structures of Solar Eruptions I: Magnetic Flux Rope (Invited Review)}

\author{X. Cheng$^{1,2}$, Y. Guo$^{1,2}$, \& M. D. Ding$^{1,2}$}
\affil{$^1$School of Astronomy and Space Science, Nanjing University, Nanjing 210093, China}
\affil{$^2$Key Laboratory for Modern Astronomy and Astrophysics (Nanjing University), Ministry of Education, Nanjing 210023, China}\email{xincheng@nju.edu.cn}

\begin{abstract}
Coronal mass ejections (CMEs) and solar flares are the large-scale and most energetic eruptive phenomena in our solar system and able to release a large quantity of plasma and magnetic flux from the solar atmosphere into the solar wind. When these high-speed magnetized plasmas along with the energetic particles arrive at the Earth, they may interact with the magnetosphere and ionosphere, and seriously affect the safety of human high-tech activities in outer space. The travel time of a CME to 1 AU is about 1-3 days, while energetic particles from the eruptions arrive even earlier. An efficient forecast of these phenomena therefore requires a clear detection of CMEs/flares at the stage as early as possible. To estimate the possibility of an eruption leading to a CME/flare, we need to elucidate some fundamental but elusive processes including in particular the origin and structures of CMEs/flares. Understanding these processes can not only improve the prediction of the occurrence of CMEs/flares and their effects on geospace and the heliosphere but also help understand the mass ejections and flares on other solar-type stars. The main purpose of this review is to address the origin and early structures of CMEs/flares, from multi-wavelength observational perspective. First of all, we start with the ongoing debate of whether the pre-eruptive configuration, i.e., a helical magnetic flux rope (MFR), of CMEs/flares exists before the eruption and then emphatically introduce observational manifestations of the MFR. Secondly, we elaborate on the possible formation mechanisms of the MFR through distinct ways. Thirdly, we discuss the initiation of the MFR and associated dynamics during its evolution toward the CME/flare. Finally, we come to some conclusions and put forward some prospects in the future.
\end{abstract}


\clearpage
\section{Introduction}
Solar eruptions refer to various phenomena that involve an outflow of plasma and magnetic flux from the solar atmosphere into the solar wind such as spicules, jets, surges, coronal mass ejections (CMEs), and flares etc.. Among them, CMEs and flares are the large-scale eruptive and energetic processes that usually accompany with each other though not always \citep{sheeley83,kahler92,yashiro06}. After shot out, CMEs often display a typical three-component structure: a leading front followed by an enclosed dark cavity and an embedded bright core \citep{illing83} as seen in white-light coronagraphs. The dark cavity and bright core are believed to be  manifestations of a magnetic flux rope (MFR), which is defined as a coherently helical magnetic structure with all field lines wrapping around the central axis at least one turn. The dark cavity may correspond to the cross section of the MFR and the bright core to the cool filament/prominence materials located at the bottom of the MFR when viewed edge-on \citep{dere99,gibson06_apj,riley08}. 

Besides being filled with a helical magnetic structure, CMEs also experience an acceleration process of short period \citep[$\sim$tens of minutes;][]{zhang01,zhang06}, finally reaching high velocities ranging from hundreds to thousands of km s$^{-1}$ \citep{yashiro04,tian12,fengli13}. After 1--3 days, these high-speed helical plasmoids may arrive at the Earth \citep{liuying11,liuying13,liuying17,shencl12_np,hess15,shitong15,huhuidong16,temmer17}. The typical features of CMEs in the interplanetary space, such as rotation of magnetic field, increased solar wind speed, depressed proton temperature, and low plasma beta, can be observed directly via in situ instruments \citep[e.g.,][]{Burlaga88,Lepping90}. When the interplanetary CMEs interact with the magnetosphere and ionosphere, they probably give rise to serious influences on the safety of human high-tech activities in outer space, such as disrupting communications, overloading power grids, presenting a hazard to astronauts, and so on \citep{gosling93,webb94,shen13,solanki04,liuying14,shitong15}.

In order to predict the products and their influences induced by solar eruptions, elucidating some fundamental but elusive processes including their origin and structures and subsequent Sun-to-Earth propagation is a matter of great importance. In the past decades, many significant progresses have been made in this aspect, the reader can refer to many previous reviews \citep[e.g.,][]{forbes06,chen11_review,schmieder15,linjun15,wangjx16,byrne10,lugaz15,mostl17}. In the current review, we elaborate on recent progresses on the study of the origin and structures of CMEs/flares from multi-wavelength observational perspective, which are mostly ascribed to the launch of Solar Dynamics Observatory \citep[SDO;][]{pesnell12}. We also introduce some relevant results from Solar Terrestrial Relations Observatory \citep[STEREO;][]{kaiser08}, Interface Region Imaging Spectrograph \citep[IRIS;][]{depontieu14}, and newly constructed ground-based instruments like the New Solar Telescope \citep[NST;][]{cao10} at Big Bear Solar Observatory and the New Vacuum Solar Telescope \citep[NVST;][]{liuzhong14} at Yunnan Observatory (Fuxian Lake). First of all, we start with the questions of whether a highly helical MFR is necessary for the eruption and whether the MFR exists prior to the eruption. We then emphatically introduce the observational manifestations of the MFR. Secondly, we elaborate on the possible formation mechanisms of the different manifestations of the MFR in Section 3. Thirdly, we discuss the initiation mechanisms of the MFR and the dynamics during the evolution of the MFR toward the CME/flare in Section 4. In the end, we come to conclusions and present some prospects that should be addressed in the future.

This review is focused on the observational aspect. The magnetic modelling aspect of the origin and structures of CMEs/flares is given in another review by \citet{guoy17}.

\section{Pre-eruptive Configurations of Solar Eruptions}
%%%%%%%%Figure 1%%%%%%%%%
\begin{figure}
\center {
\includegraphics[width=17cm]{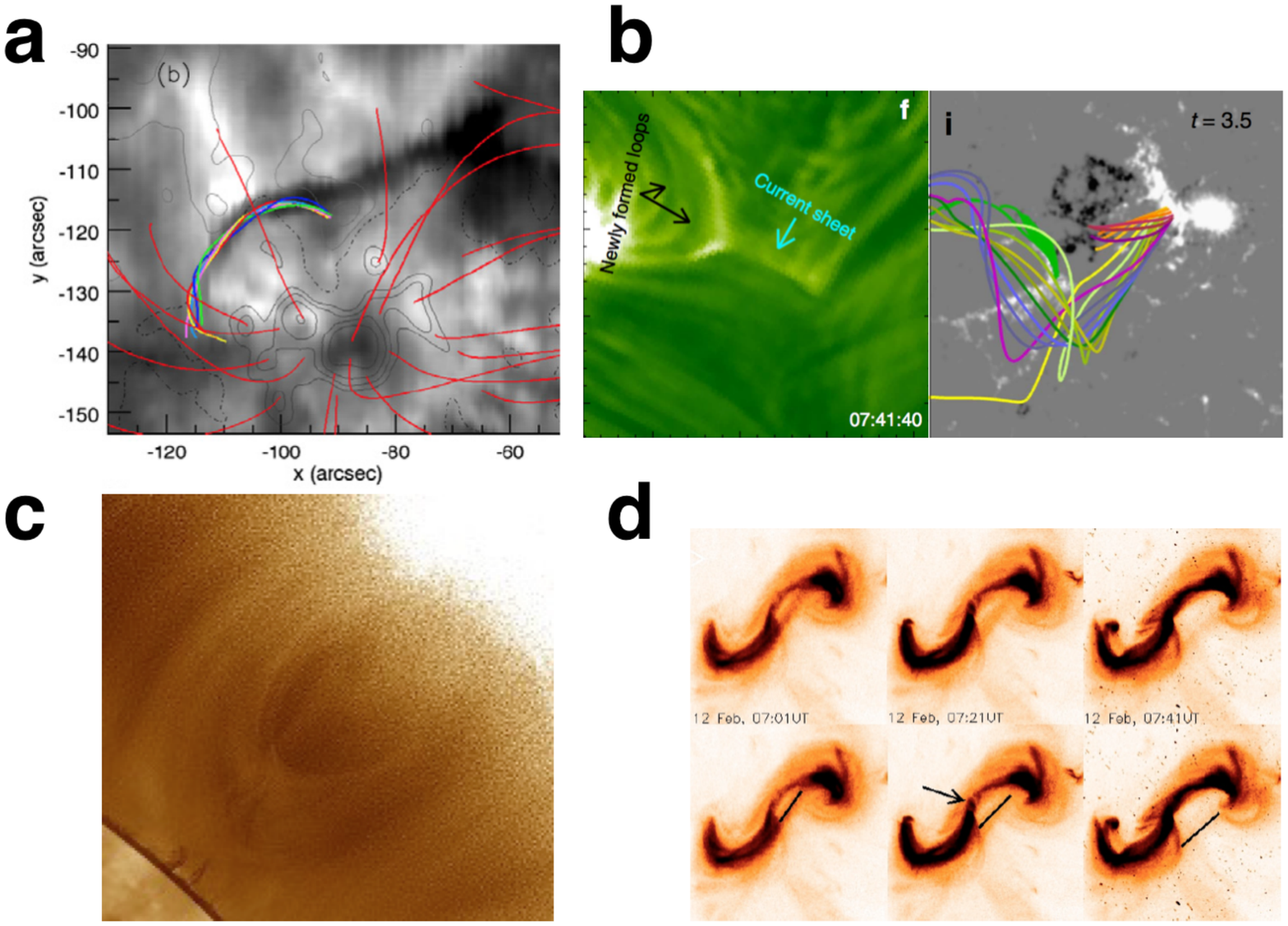}}
\vspace{-0.1\textwidth}
\caption{(a) H$\alpha$ image overlaid by the extrapolated magnetic field lines. The MFR indicated by mixed colors is cospatial with a segment of the filament \citep[adapted from][]{guo10_filament}. (b) Twist releasing by the reconnection during the filament eruption \citep[for details please see][]{xuezhike16}. (c) Coronal cavities as seen in the AIA 193 {\AA} passband. (d) A sequence of XRT images showing the evolution of a sigmoid prior to the eruption \citep[from][]{mckenzie08}.}
\label{evidence}
\end{figure}
%%%%%%%%Figure 1%%%%%%%%%

In the 2D standard CME/flare model \citep{carmichael64,sturrock66,hirayama74,kopp76,shibata95}, the pre-eruptive configuration, which is modelled to be a helical MFR \citep{shibata95,chenj96,titov99,vourlidas13} or sheared arcade \citep{sturrock66,antiochos99}, is constrained by the background magnetic fields. The eruption of the pre-eruptive configuration stretches the background fields to form a magnetic dissipation region, a so-called current sheet (CS), in between their two legs. Once the thickness of the CS is less than a threshold, magnetic reconnection will be switched on \citep{linjun00}. On the one hand, the reconnection accelerates the eruption via continuously injecting poloidal flux into the erupting structure. On the other hand, the reconnection releases a large quantity of energy that induces a rapidly enhanced radiation over the whole electromagnetic spectrum ranging from decameter radio waves to $\gamma$ rays. 

At present, the nature of the pre-eruptive configuration is still elusive. On the one hand, observations show that the pre-eruptive configuration could be sheared arcades, indicating that the MFR could be unnecessary for initiating the eruption \citep[e.g.,][]{song14_formaion,ouyang15}. On the other hand, a few observations imply that the pre-eruptive configuration is a helical MFR \citep[e.g.,][]{low95_apj,gibson06_ssr,green09,zhang12,cheng13_driver,patsourakos13}. Considering that the magnetic field in the solar atmosphere cannot be measured accurately except on the photosphere, the community usually resorts to indirect observations or extrapolation techniques such as non-linear force-free field (NLFFF) modelling to search for the evidence of the MFR. For example, through the NLFFF modelling, strongly twisted field lines with a substantial magnetic helicity are often found to exist along the polarity inversion line (PIL) of active regions before the eruption \citep[e.g.,][]{yanyh01,canou10,guo10_filament,savcheva12a,cheng13_double,cheng14_formation,inoue13,jiang14_nlfff,jiang16a,yanxl15}. In the following, we introduce various observational evidence for the existence of the MFR in detail.

\subsection{Filaments and Filament Channels}
Filaments are a phenomenon of relatively cool and dense plasma embedded in the hot and tenuous corona, commonly observed in absorption in H$\alpha$ and the Extreme Ultraviolet (EUV) passbands on the solar disk, while appearing in emission as bright features, i.e., prominences, against the dark background when seen above the solar limb \citep{hirayama85,mackay10}. 

The magnetic structure of filaments is usually thought to be sheared arcades \citep{antiochos94,aulanier06} or highly twisted MFR \citep{kuperus74,vanballegooijen89,aulanier98,aulanier99}, which possess magnetic dips that are able to provide an upward magnetic tension against the gravity of filament materials \citep{martin98,mackay10}. In order to validate such a picture, many authors extrapolated three-dimensional (3D) structures of the filament source regions based on the assumption of linear \citep[e.g.,][]{aulanier98,aulanier99} or non-linear FFF \citep[e.g.,][]{guo10_filament,savcheva12a,cheng14_formation,jiang16a,yanxl15}. In many events, in particular active region filaments, the dips of the sheared arcades \citep{antiochos94,aulanier06} or twisted fields \citep[e.g.,][]{cheng14_formation,jiang16a} are mostly co-spatial with the filament locations. Sometimes, a part of the filament locations are consistent with the dips of sheared arcades, while the other part with the dips of twisted fields \citep[Figure \ref{evidence}a,][]{guo10_filament}. During the filament eruption, the twist of magnetic field lines is also observed to be released by magnetic reconnection \citep[Figure \ref{evidence}b;][]{xuezhike16}.

In some events, it is very difficult to reconstruct the strongly twisted field lines comparable with the filaments, in particular for the quiescent filaments, which is probably due to the reason that the preprocessing over-smoothes the vector field before doing the extrapolation. However, using the newly developed CESE--MHD--NLFFF code by \cite{jiang12}, \cite{jiang14_filament} reproduced a large-scale coronal MFR that exists in between an active region and a weak polarity region and supports a quiescent filament. It is even found that the large polar crown prominence located at the weak magnetic field region can also be modelled with an MFR configuration although with a certain degrees of freedom \citep[e.g.,][]{suyingna12,suyingna15}. 

It is possible that there are no cool materials deposited in dips of MFRs. In this case, the MFRs may manifest as filament channels and usually lie over the PIL of the long decayed active regions \citep[e.g.,][]{vanballegooijen98,aulanier02,chenpf14}. It is also observed that the erupting MFR does not accompany with a pre-existing filament, such as for a double-decker configuration that consists of a high-lying MFR and a vertically separated filament-associated low-lying flux system \citep[e.g.,][]{liurui12_filament,cheng14_formation,dudik14,kliem14}. With the eruption beginning, only the high-lying MFR flux erupts to give rise to a CME and a flare, while the low-lying filament remains in original place.

\subsection{Coronal Cavities}
When quiescent filaments and filament channels rotate to the solar limb, they are probably seen as dark, semi-circular or circular cavities surrounding prominences and embedded in bipolar helmet streamer (Figure \ref{evidence}c). Cavities in active regions are very difficult to observe, because they lie relatively low to the solar surface and are significantly influenced by strong emission from the foreground and background. At present, they have been observed only in few events as an erupting hot blob \citep[e.g.,][]{song14}. It is also argued that the magnetic structure of cavities is an MFR, i.e., the cross section of the MFR corresponds to the whole or lower part of cavities \citep{low95_apj}. Prior to the eruption, the cavities typically exist in the low corona and are able to survive for days, even for months \citep{gibson06_ssr,gibson06_apj}. It can be observed at a range of wavelengths, mostly in the white-light passband of such as the ground-based white-light coronagraph Mark IV coronameter installed at the Mauna Loa Solar Observatory, as well in the EUV passbands of such as the Atmospheric Imaging Assembly \citep[AIA;][]{lemen12} on board SDO.

Many features of coronal cavities indicate that their fundamental magnetic structure is an MFR. The first evidence is continuous spinning motions, which are frequently seen inside cavities and have a flow speed of 5-10 km s$^{-1}$ \citep{wangym10}. Moreover, the polarization ring in cavities observed by Coronal Multi-Channel Polarimeter also supports the MFR model, which illustrates that a bright ring of linear polarization may appear in a density depleted region \citep{dove11}. In linear polarization observations, \cite{bak-steslicka13} further found that the cavity possesses a characteristic ``lagomorphic" signature, which again indicates the existence of the MFR as a pattern of concentric rings.

Solar ``tornadoes'', a new phenomenon discovered recently and often appearing in cavity-associated prominences, are also considered as a piece of evidence of the MFR. The direct evidence of `tornadoes'' having a helical structure is the swirling motions \citep{zhangjun11,lixing12,suyang12_tornadoes,wedemeyer12}. Spectroscopic observations also disclosed a pattern with blueshifted and redshifted emissions, i.e., opposite velocities, existing at the two sides of prominences, implying the magnetic structure of ``tornadoes'' being helical \citep{suyang14_tornadoes}. However, spectroscopic observations in cool lines (e.g., H$\alpha$ and 10830 {\AA}) revealed that the Doppler shift pattern does not follow the pattern observed in coronal lines. It is most likely oscillations of the plasma along the field lines like counterstreaming or oscillations of the whole magnetic structure \citep{martinez16,schmieder17}. Tornadoes could be just the foopoints of prominences \citep{wedemeyer13,levens16} or a manifestation of spirally ejected jets driven by torsional Alfv$\acute{e}$n waves \citep{pariat09_jet}.

\subsection{Sigmoids}
Sigmoids, forward or reversed sigmoidal emission patterns appearing in EUV and soft X-ray (SXR) passbands (Figure \ref{evidence}d), have been found to be an important pre-eruptive configuration of CMEs/flares \citep{hudson98,rust96,sterling97,gibson02}, which are statistically more likely to be eruptive \citep{canfield99}. Based on the duration time, sigmoids can be classified as transient or persistent ones. The former tend to be sharper and brighter, apparently as sigmoidal loops, and evolve into cusps or arcades of loops many times; the latter appear more diffuse and could be a collection of some sheared loops \citep{pevtsov02,gibson02,green07}. The sigmoidal emission pattern is expected to be due to the heating in a curved CS at the interface between the helical core field (e.g., MFR) and the ambient field \citep{kliem04,gibson06_ssr}. \cite{green07} even found that, during the eruption phase of sigmoids, the helicity sign of sigmoids is consistent with the rotation direction of associated erupting filaments \citep[also see][]{yangsb15}, showing the conversion of twist into writhe under the assumption of helicity conservation, supporting the existence of the twisted field lines in sigmoids.

The appearance of the sigmoidal emission pattern does not mean the existence of continuous sigmoidal field lines. \cite{mckenzie08} analysed a long-lasting coronal sigmoid and found that the overall S shape of the sigmoid definitely consists of two separate J-shaped loops with a straight section possibly lying in the middle. \cite{green09} and \cite{liur10} pointed out that two opposite J-shaped loops can form the continuous S-shaped loops through the tether-cutting reconnection. Using an MHD simulation, \cite{aulanier10} reproduced synthetic SXR images from the distribution of the electric currents and revealed the formation of a sigmoidal active region. They found that a bright sigmoidal envelope is built up gradually by the bald-patch (BP, where magnetic field lines are curved upward and are tangent to the photosphere) and tether-cutting reconnection between two pair of J-shaped field lines. Using the flux emergence model, \citet{archontis09_sigmoid} even disclosed that opposite J-shaped loops and S-shaped loops exist simultaneously, which result in the overall magnetic structure of sigmoids. Moreover, some authors also reconstructed 3D NLFFF configuration of source regions of sigmoids and did find that the core field consists of a twisted MFR embedded in highly sheared fields \citep[e.g.,][]{suyingna09,savcheva09,savcheva12a,jiang13,jiang14_nlfff,cheng14_formation}.

%%%%%%%%%%Figure 2%%%%%%%%%%%%%%
\begin{figure}
\center {
\includegraphics[width=17cm]{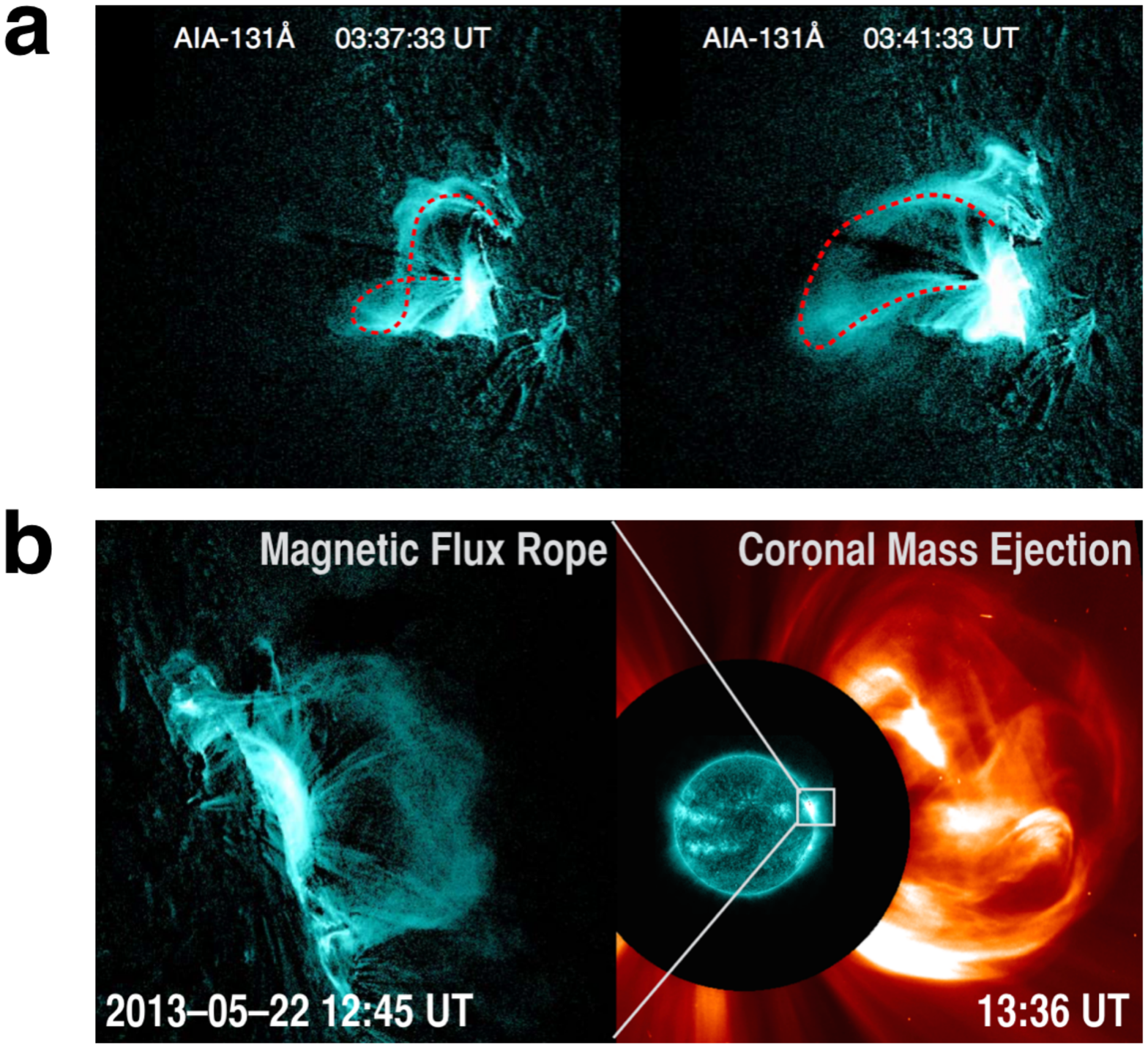}}
\caption{(a) AIA 131 {\AA} images showing the pre-existence (left) and eruption (right) of a hot channel-like MFR \citep[from][]{zhang12,cheng13_driver}. (b) Transformation of a hot channel-like MFR, as seen in the AIA 131 {\AA} passband (left), to the CME imaged by the LASCO C2 white-light coronagraph (right).}
\label{channel}
\end{figure}
%%%%%%%%%%Figure 2%%%%%%%%%%%%%%

\subsection{Hot Channels}
Hot channels or hot blobs are a type of new and promising evidence of the existence of MFRs. Through analysing a limb event, \citet{cheng11_fluxrope} for the first time observed the formation of an MFR during the impulsive phase. It initially appears as an erupting hot blob as seen in the AIA 131 {\AA} and 94 {\AA} passbands (T$\ge$8 MK). While in the other low temperature passbands (1MK$\le$ T $\le$5 MK), it appears as a dark cavity. Combing with some typical features such as the inflows, stretched overlying field, and cusp-shaped flare loops that are expected by the MFR eruption models of CME/flares, the authors strongly argued that the hot blob is an unambiguous evidence of the MFR existing in the corona \citep[also see;][]{song14}. Following the work of \citet{cheng11_fluxrope}, \citet{zhang12} and \citet{cheng13_driver} started to search for more evidence of the MFR in the 131 {\AA} and 94 {\AA} passbands. They discovered that the MFR even exists prior to the eruption as a writhed channel-like structure with two elbows inclining to the opposite directions and the middle being concaved toward the surface when seen off the solar limb (Figure \ref{channel}a). The visibility of the channel-like structure only at the AIA high temperature passbands (e.g., 131 {\AA} and 94 {\AA}) but not at other cooler passbands shows that it has a temperature of $>$6 MK. Subsequently, more and more hot channel events that exist prior to the CME/flare beginning are identified \citep[e.g.,][]{patsourakos13,lileping13,liting13_homologous,tripathi13,vemareddy14,dudik14,chintzoglou15,joshi15,zhougp16}. Interestingly, a pre-existing MFR is even confirmed to exist in the chromosphere with observations by NST at BBSO \citep{wanghm15_nc}. Moreover, \citet{cheng12_dem} quantified the differential emission measure (DEM) of the hot blobs and channels, which shows that the emission of these hot MFRs are actually from a broad temperature range of 6.5 $\leq\log T \leq$ 7.3 with a DEM-weighted average temperature larger than $\sim$8 MK. The corresponding electron number density varies from 5.0 $\times10^{8}$ to 3.0 $\times10^{9}$ cm$^{-3}$.

\citet{zhang12} and \citet{cheng13_driver} further found that the hot channel shows a remarkable morphological evolution during the early phase of the eruption. Initially, the dipped central part of the writhed hot channel rises up slowly and gradually becomes more linear. The continuing rise of the central part eventually turns the sigmoidal shape of the channel into a loop-like shaped partial torus (Figure \ref{channel}a). During the transformation process, the two footpoints of the evolving hot channel are nearly fixed. Afterwards, the loop-like structure quickly stretches the overlying field and builds up a CME, simultaneously giving rise to a flare underneath. Furthermore, \citet{cheng14_tracking} identified that the hot channel is capable of evolving smoothly from the inner into the outer corona with almost retaining its coherence, morphologically consistent with the CME cavity as seen in the white-light images (Figure \ref{channel}b). \cite{cheng16_apjs} studied the footpoints of the hot channel and found a substantial deviation of the hot channel axis from the associated filament. It shows that the hot channel has ascended to a high altitude and likely separated from that of the filament when approaching the eruption. In order to uncover the appearance frequency of the hot MFR, \citet{nindos15} made a statistical study and documented that almost half of major eruptive flares contain a hot blob or channel-like configuration. The observed MFR morphology mainly depends on the orientation of the MFR axis with respect to the line of sight. That is to say, the MFR appears as a hot blob and a hot channel parallel and perpendicular to its axis, respectively.

In addition, some spectroscopic observations also support the pre-existence of the MFR. By analysing Coronal Diagnostics Spectrometer (CDS) or EUV Imaging Spectrometer (EIS) data, \citet{gibson02}, \citet{harra09}, and \citet{harra13} found a significant pre-flare enhancement in non-thermal velocity, the locations of which may correspond to the footpoints of the MFR. Combining EIS and AIA observations, \citet{syntelis16} even found that the enhanced non-thermal velocities, as well as the blueshifts, can last for 5 hours before the eruption of the hot channel-like MFR.

\subsection{Reconciling Distinct Aspects of the MFR}
As discussed above, filaments, filament channels, cavities, sigmoids, and hot channels can be well unified in the framework of the MFR; they may be just the distinct manifestations of the MFR, depending on different observational wavelengths and perspectives, as well as magnetic environment. There have been many studies revealing the relationship between any two of them.

Through observing the evolution of an erupting filament, \citet{liting13} found that the erupting materials have a helical trajectory when moving along the threads of hot channels \citep[also see][]{yang14,zhangjun15}. During the eruption phase of hot channels, cool filamentary materials are also seen to descend spirally down to the chromosphere along their legs \citep{cheng14_tracking}. Furthermore, \citet{cheng14_kink} found that the hot channel is initially co-spatial with the prominence in the early rise phase, while with the eruption beginning the hot channel quickly expands, resulting in a separation of its top from the prominence. Through a detailed analysis of the temperature structure of an erupting filament, \citet{chenbin14} confirmed that the relatively cool plasma always stays at the bottom of the hot channel. These results strongly suggest that the hot channel is a direct manifestation of the heated MFR with filament materials collected at its bottom.

Although previous studies revealed that CME-productive active regions often take on a sigmoidal shape in the EUV and/or SXR images prior to the eruption, it does not mean that the corresponding magnetic field lines must be highly twisted. Alternatively, they could consist of two groups of sheared arcades, making up a sigmoidal shape apparently \citep{titov99,kliem04,schmieder15,cheng16_apjs}. However, we still cannot exclude the possibility that an existing but invisible MFR, probably having a very weak emission, is embedded in the middle of the sigmoid and overlaid by ambient sheared arcades. Recently, people have started to recognise that the continuous sigmoidal or highly twisted field lines can originate in the sigmoidal active regions. Using XRT data, \cite{mckenzie08} observed a diffuse linear structure that appears in the middle of the sigmoid prior to the eruption and lifts off as the flare begins \citep[also see][]{liur10,green11,zharkov11}. Taking advantage of the unprecedented high cadence, high resolution, and multi-wavelength observations of the AIA, \cite{cheng14_formation} found that the linear feature is most likely to be continuous sigmoidal hot threads. They even found that a double-decker MFR system that consists of a high-lying continuous sigmoidal threads (hot channel) and a vertically separated filament-associated low-lying flux could be formed in the sigmoidal active region. Close to the eruption, the morphology of the high-lying hot channel varies from an S-shape to a loop-shape, similarly to the linear feature in the erupting sigmoids. In addition, it should be noted that sigmoids are mostly a manifestation due to the heating in a sigmoidal CS between the MFR and its ambient field \citep{kliem04,gibson06_ssr}, while they do not delineate the specific magnetic field configurations. However, hot channels refer to coherent magnetic structures, which can be traced continuously from the sigmoidal active regions to the outer corona. During the whole eruption process, the evolution of hot channels is mainly controlled by their own dynamics. Therefore, we can say that hot channels and sigmoids are distinct phenomena, more specifically, the former are well-defined and specific structures that originate in the latter. 

The visibility of hot channel-like MFRs only in the 131 {\AA} and 94 {\AA} passbands shows that they are substantially heated before the eruption. Interestingly, quiescent cavities are also found to be heated with a higher temperature than the background. \citet{reeves12} examined the thermal properties of a quiescent cavity that contains strong X-ray emission in its core and found that there is an obvious temperature increase in the cavity core, and that the core temperature varies from 1.75 MK to 2.0 MK with the evolution of the morphology from a ring-shaped at the beginning to an elongated structure two days later. The reason is conjectured to be that different parts of the cavity core are heated at different times. By constructing limb synoptic maps of the AIA 211 {\AA}, 193 {\AA}, and 171 {\AA} passbands, \citet{karna15} analysed a number of quiescent cavity events and also found that quiescent cavities are hotter than their surroundings although only slightly. These results imply that active region hot channels and quiescent cavities may have the same, at least similar, heating mechanism in the pre-eruptive phase, though the exact mechanism remains mysterious at present. We believe that magnetic structures of both the active region hot channels and quiescent cavities are an MFR, the only difference of them is the distinct size; the former usually has a length of 20--100 Mm (the scale of the active region PIL) and a height of 10--20 Mm, while the latter extends along the whole PIL of long-term decayed active regions, having a length of 200--500 Mm and a height of 30--100 Mm \citep[e.g.,][]{liur10,suyingna15,cheng16_apjs}. 

\section{Formation of Pre-eruptive Configurations}

%%%%%%%%%%Figure 3%%%%%%%%%%%%%
\begin{figure}
\center {\vspace{-0.25\textwidth}
\includegraphics[width=15cm]{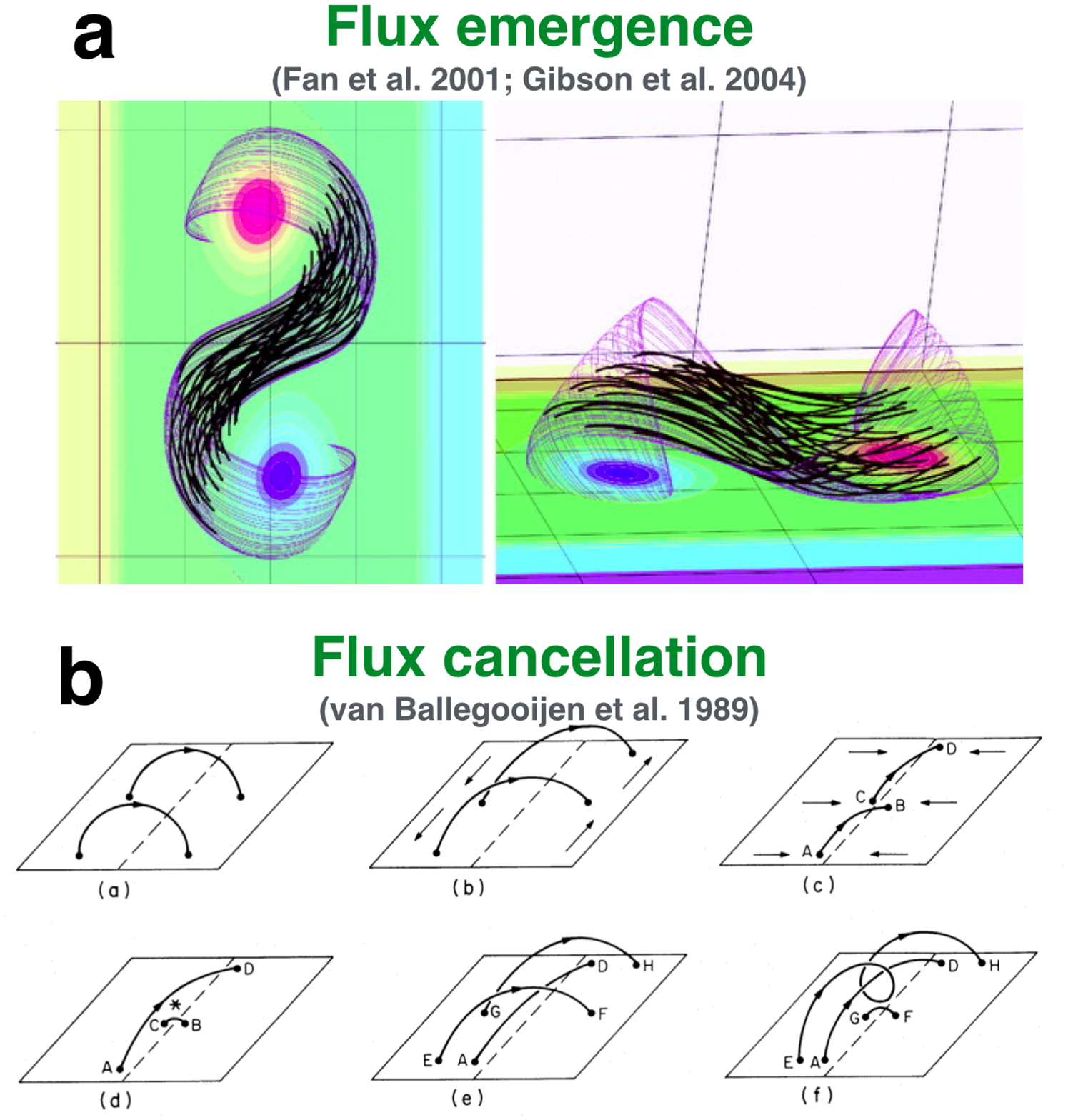}}
\vspace{-0.17\textwidth}
\caption{(a) Flux emergence model, in which a twisted MFR bodily emerges from below the photosphere to the corona. The lines in violet show bald-patch--associated separatrix surfaces. The black segments display magnetic dips where the filament materials can be collected \citep[for details please see][]{gibson04}. (b) Flux cancellation model, in which the MFR is formed by the reconnection of two group of sheared arcades driven by the shearing and converging motions \citep[for details please see][]{vanballegooijen89}.}
\label{formation}
\end{figure}
%%%%%%%%%%Figure 3%%%%%%%%%%%%%

\subsection{Bodily Emergence of the MFR}
If an MFR really exists in the corona, the question is then when, where, and how the MFR is built up. Theoretically, two possibilities have been proposed. One possibility is that the MFR in the convection zone emerges into the corona by buoyancy \citep[Figure \ref{formation}a;][]{fan01,magara04,Martnez-sykora08,archontis08_aa,leake13}. However, \citet{manchester04} found that when the primary axis of the MFR approaches the photosphere, the MFR is split into two parts by magnetic reconnection with surrounding fields, which only allows the upper flux of the middle section with very weak twist (less than one turn about the axis) to separate from the lower mass-laden and dipped flux \citep[also see][]{magara06}. Even so, the total relative magnetic helicity of the whole system is well conserved \citep{zhangm03}. After ascending to the corona, the center of the MFR rises with an increasing velocity as long as the MFR footpoints rotate continuously. As a result, significant twist is transported from the MFR interior part toward the coronal part through nonlinear torsional Alfv$\acute{e}$n waves \citep{fan09,leake13}. After emerging into the corona, the reconnection with the pre-existing coronal field also plays an important role in forming the MFR and even driving its eruption \citep[e.g.,][]{archontis08_aa,leake14}. \citet{fan12} found that in the quasi-static rise phase of the MFR, magnetic reconnection, most likely tether-cutting in the sigmoidal hyperbolic flux tube (HFT, intersection of two quasi-separatrix layers (QSLs), where the linkages of magnetic field lines are continuous but change drastically \citep{titov02}), effectively injects twisted flux to the MFR so as to drive its eruption.

Some observational studies also stand for the emergence of the MFR from below the photosphere to the corona. Through analyzing the vector magnetograms obtained by the Dunn Solar Telescope of the National Solar Observatory, \citet{lites05} found a concave-up geometry in the photosphere below two active region filaments. \citet{okamoto08} and \citet{okamoto09} examined a sequence of vector magnetograms of AR 10953 observed with the Solar Optical Telescope on board Hinode and found the following features: the adjacent opposite-polarity regions with horizontally strong but vertically weak magnetic fields growing laterally and then narrowing, the reversal of the direction of the horizontal magnetic fields along the PIL from a normal polarity to an inverse one, and the appearance of the blueshift and diverging flows in the horizontal magnetic field region. These observational features, as well as the concave-up geometry, imply that the MFR probably may emerge from the solar interior to the corona. However, \citet{Dominguez12} recently provided a contradictory interpretation for those observational characteristics in the photosphere. Comparing with the numerical results of \citet{mactaggart10}, they pointed out that magnetic cancellation is also capable of producing the lateral growing and then narrowing of the positive and negative polarities, as well as the reversal of the direction of the horizontal magnetic fields.

Magneto-convection has a significant influence on the emergence of the MFR from below the photosphere to the corona. Because of convective flows, undulations appear in the emerging horizontal field to form $\Omega$-loops and U-loops \citep{cheung14}. The later naturally have a concave-up geometry. \citet{bernasconi02} and \citet{Pariat04} studied the Flare Genesis Telescope data and found that serpentine structures and ``U"-shaped loops frequently appeared in emerging active regions. Ellerman bombs are also detected at the locations where serpentine structures and ``U"-shaped loops touch the photosphere \citep[also see][]{lizheng15}. This is mainly due to the buildup of currents along the serpentine and ``U"-shaped magnetic field, which then lead to the reconnection in the lower atmosphere \citep[e.g.,][]{isobe07,pariat09_emerging,archontis09,wang06}. For more details concerning how the sub-photosphere magnetic field emerges into the corona and produces various actives, the reader can consult the reviews by \cite{Schmieder14} and \citet{cheung14}.

\subsection{MFR Formation by Magnetic Reconnection}
\subsubsection{MFR Formation prior to the Eruption}
The MFR can also be built up directly in the corona via magnetic reconnection prior to the eruption. In the model of \citet{vanballegooijen89}, it is proposed that flux cancellation transfers sheared loops to helical field lines, creating a coherent MFR configuration (Figure \ref{formation}b). The flux cancellation is usually interpreted in terms of transport of positive and negative fluxes toward the PIL, reminiscent of the well-known moat flow around two polarities of an active region \citep{amari10,amari11,amari14}. Through imposing converging motions toward the PIL, \citet{amari03a} successfully simulated that two groups of sheared flux are brought together and reconnect toward a twisted MFR \citep[also see][]{mackay06}. \citet{amari99,amari03b} showed the importance of photospheric turbulent diffusion on pre-sheared magnetic field (possibly remnant) that leads to the formation of magnetic flux rope. Subsequently, \citet{aulanier10} did a more detailed MHD simulation, in which an initially potential bipolar field evolves as driven by magnetic field diffusion and shearing motions. Similar to the results of \citet{amari03a}, flux-cancellation-driven reconnection appears in a BP separatrix and gradually transforms the sheared arcades into the MFR. In the whole formation process, the MFR gradually rises up but in a quasi-static manner. Then, the BP structure changes to the HFT topology, where the reconnection, of a tether-cutting type, takes place to continuously inject the poloidal flux to the MFR. Using isothermal MHD simulations, \citet{xia14} evolved a linear force-free bipolar magnetic field by means of introducing vortex flows around the opposite polarities and converging flows toward the PIL. They also found the creation of the helical field lines through the reconnection and flux cancellation at the PIL driven by the converging flows.

Observationally, a direct view of the formation of an MFR is impossible as magnetic field measurement above the photosphere is technically difficult at present. Thus, for the sake of exploring the formation of the MFR, people usually investigate how the various manifestations of the MFR, including filaments, sigmoids, and hot channels, are formed. 

%%%%%%%%%%Figure 4%%%%%%%%%%%%%
\begin{figure}
\center {\hspace{-0.1\textwidth}
\includegraphics[width=18cm]{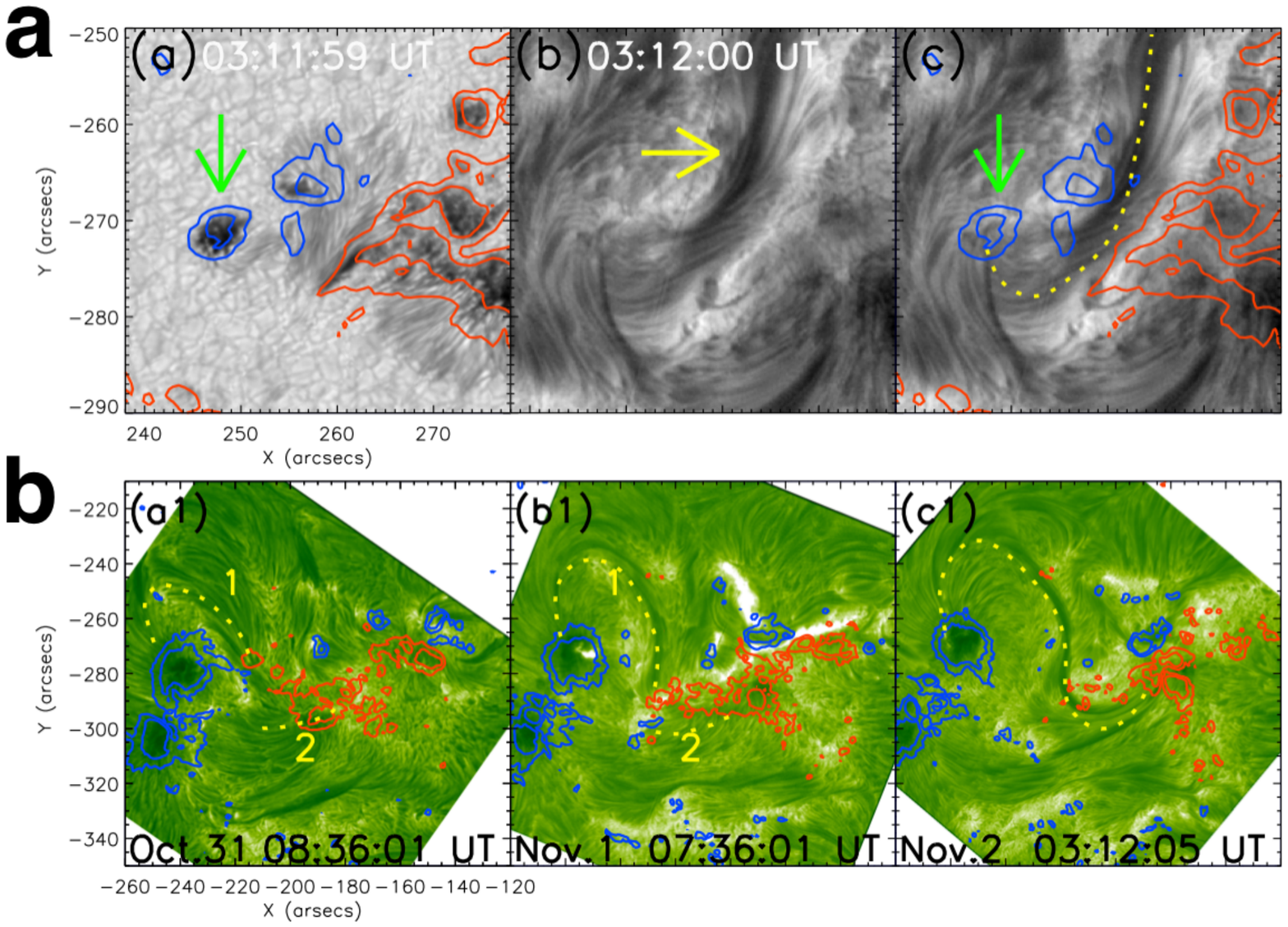}}
\vspace{-0.15\textwidth}
\caption{(a) NVST TiO and H$\alpha$ images overlaid by line-of-sight magnetograms with the positive (negative) in blue (red) showing the formation of a filament driven by the sunspot rotation \citep[from][]{yanxl15}. (b) NVST H$\alpha$ images displaying the formation of a filament by the reconnection \citep[from][]{yanxl16}.}
\label{filament_formation}
\end{figure}
%%%%%%%%%%Figure 4%%%%%%%%%%%%%

Through H$\alpha$ observations by the Multi-channel Subtractive Double Pass spectrograph, \cite{schmieder04} observed that different segments of filaments merge (reconnect) to form a long filament. At the mergence locations, both brightenings at the EUV passbands and flux cancellations of small bipolar are found. This is consistent with the filament formation model in terms of magnetic reconnection proposed by \citet{vanballegooijen89} and \citet{aulanier98}. After the reconnection, the cool materials are expelled along the reconnected field lines, which is confirmed by measured horizontal velocities \citep[e.g.,][]{deng02}.

Recently, using H$\alpha$ data with higher resolution and cadence provided by the NVST, \citet{yanxl15} observed the obvious shearing motion of the opposite polarities and the sunspot rotation during the formation process of two active-region filaments (Figure \ref{filament_formation}a). They suggested that the shearing motion stretches filament-associated magnetic field more horizontal and then the sunspot rotation injects some twist to form a filament-hosting helical magnetic structure. Besides the sunspot rotation, \citet{yanxl16} and \citet{vemareddy16} also addressed the role of the reconnection in building up the helical configuration, which is evidenced by the appearance of the EUV/UV brightening at the touch point of the different branches (Figure \ref{filament_formation}b). By means of studying the interaction of two sets of dark threads or filament channels driven by flux convergence and cancellation, both \citet{joshi14_filament} and \citet{yangbo16} argued that the reconnection is a necessary condition for the formation of the filament. Moreover, they also observed that the reconnection-driven hot plasma undergo a rolling motion along the filament threads.

\citet{tripathi09} analysed the temperature structure of a sigmoid and discovered that the plasma in the J-shaped arcades has a higher temperature than that in the S-shaped flux if both are simultaneously visible. They argued that the J-shaped arcades are most likely reconnecting to the S-shaped flux, thus having a higher temperature but starting to cool down after leaving the reconnection diffusion region. \citet{green09} and \citet{green11} supported the point that the sigmoid is from the reconnection of sheared arcades that is driven by the flux convergence and cancellation under the sigmoid although only part of the cancelled flux being injected into the sigmoidal field lines. 

%%%%%%%%%%Figure 5%%%%%%%%%%%%%
\begin{figure}
\center {\hspace{-0.06\textwidth}
\includegraphics[width=15cm]{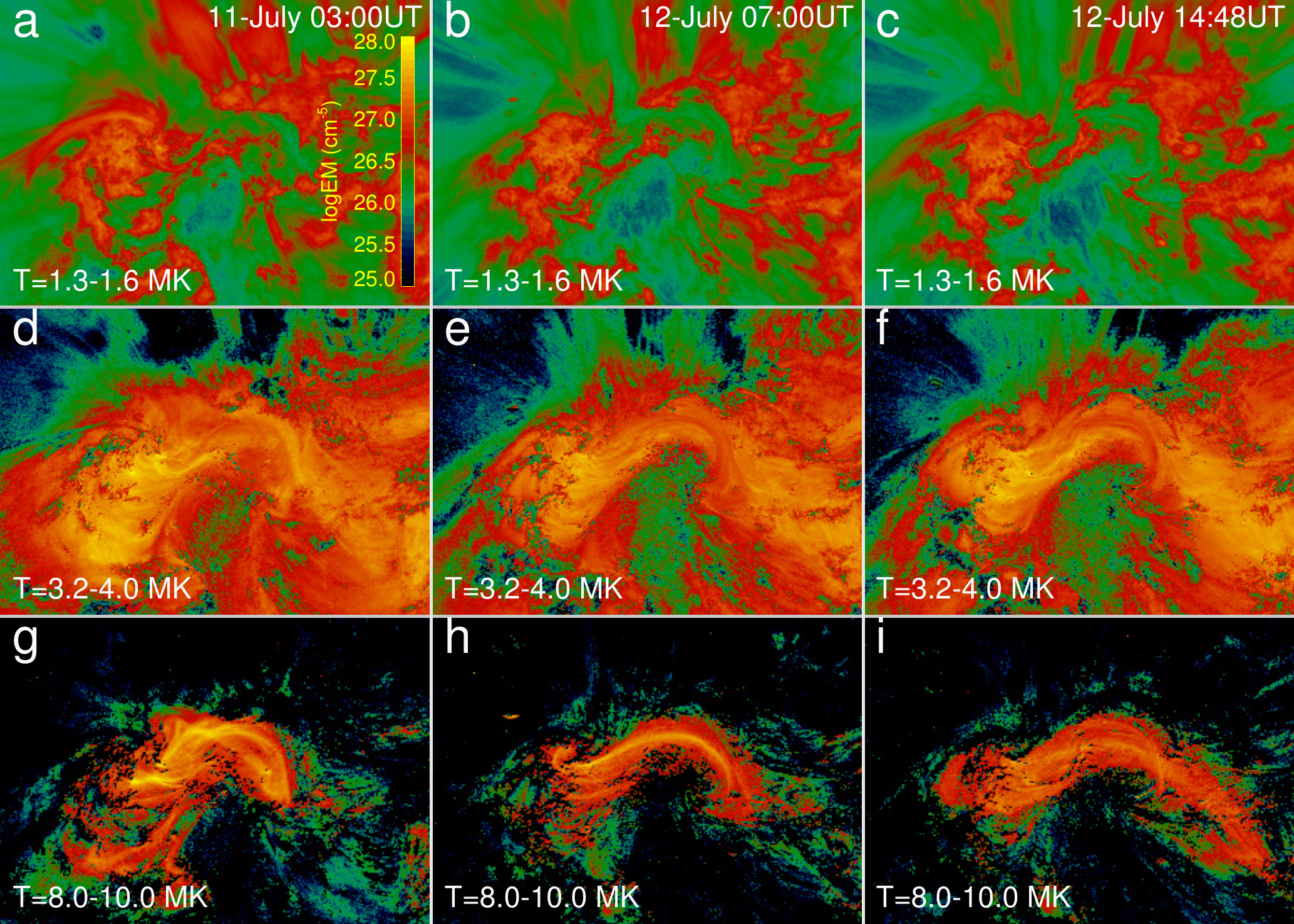}}
\caption{Emission measure maps at different temperature intervals and instants showing the formation of a sigmoidal hot channel-like MFR in the active region 11520, which can be clearly seen in panels g--i \citep[adapted from][]{cheng14_formation}.}
\label{dem}
\end{figure}
%%%%%%%%%%Figure 5%%%%%%%%%%%%%
\citet{cheng14_formation} studied the formation of the hot channel-like MFR after that was discovered. Through analysing the long-term evolution of an evolving sigmoidal active region, they found that the twisted field, indicated by continuous sigmoidal hot threads, is formed via the reconnection of two groups of sheared arcades near the PIL half day before the eruption. The temperature of the twisted field and sheared arcades derived by the DEM technique is higher than that of the ambient volume (Figure \ref{dem}a), indicating that the reconnection takes place and heat the plasma therein. They also confirmed that the reconnection is driven by the shearing and converging motions near the PIL. Through constructing a time sequence of NLFFF structures, it is further revealed that the reconnection happens simultaneously at the BP separatrix in the photosphere and in the HFT in the corona (the tether-cutting). The MFR can even be formed in the lower atmosphere \citep[e.g.,][]{wanghm15_nc}, via, for example, a series of magnetic reconnection in the chromosphere, and sometimes be heated up to the coronal temperature as visible in the AIA 131 {\AA} and 94 {\AA} passbands \citep{kumar15,kumar17,liting15_solarphy}. Moreover, the conversion of mutual helicity to self-helicity through the interchange reconnection of two group of loops is also argued to be strong evidence for the formation of the helical field prior to the eruption \citep[e.g.,][]{tziotziou13,lileping14}.

\begin{figure}
\center {\hspace{-0.02\textwidth}
\includegraphics[width=16cm]{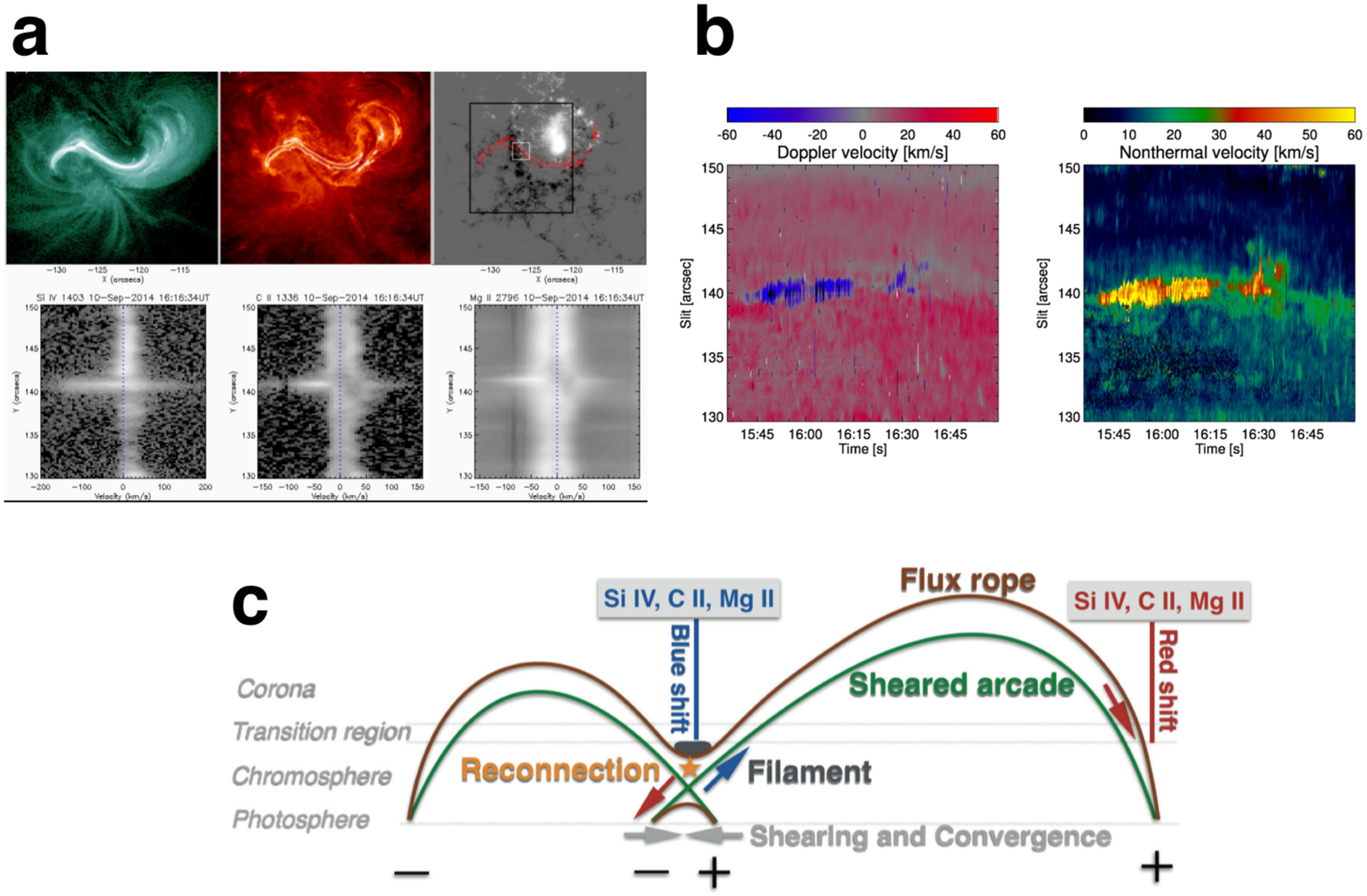}}
\vspace{-0.04\textwidth}
\caption{(a) SDO/AIA 131~{\AA}, 304~{\AA} images, and \textit{SDO}/HMI line-of-sight magnetogram showing the formation of an MFR prior to the eruption (top). Spectrograms of the \ion{Si}{4}, \ion{C}{2}, and \ion{Mg}{2} lines at the MFR formation site (bottom). (b) Doppler velocity and non-thermal velocity maps of the \ion{Si}{4} line at the MFR formation site. (c) A cartoon illustrating the MFR (brown) formation through the reconnection of two arcades (green) in the lower atmosphere \citep[for details please see][]{cheng15_iris}.}
\label{iris}
\end{figure}

\citet{cheng15_iris} further performed spectroscopic diagnostics on the formation of hot channels based on the AIA and \textit{IRIS} joint observations. At the footpoints of the hot channel, it is found that the \ion{Si}{4}, \ion{C}{2}, and \ion{Mg}{2} lines exhibit weak to moderate redshifts and non-thermal velocities in the pre-flare phase. However, relatively large blueshifts and extremely strong non-thermal velocities appear at the reconnection site of two sheared arcades, i.e., the formation site of the hot channel (Figure \ref{iris}a and \ref{iris}b). These spectral features imply that the reconnection plays an important role in the formation and heating of hot channels, and that the location of the reconnection is most likely in the lower atmosphere (Figure \ref{iris}c), based on the fact that the \ion{Si}{4}, \ion{C}{2}, and \ion{Mg}{2} lines, forming in the chromosphere and transition region, all exhibit blueshifts and non-thermal velocities. The outflows from the reconnection site may propagate toward the footpoints of the hot channel along the newly reconnected field lines, producing weak redshifts and non-thermal velocities. Note that, redshifts are also expected at the reconnection site \citep{innes97,peter14}, which, however, could be absent in the observed lines because of kinetically being less obvious than blueshifts. We should also mention that, the reconnection is not a unique interpretation for the appearance of blueshifts, redshifts, and non-thermal velocities. The rotation motion could be an alternative reason.

%%%%%%%%%%Figure 6%%%%%%%%%%%%%
%%%%%%%%%%Figure 6%%%%%%%%%%%%%

It is worth noting that the MFR can even be formed during a series of confined flares prior to the eruption. \citet{patsourakos13} studied a confined flare and an eruptive flare from the same source region and believed that the first confined flare forms the MFR by the reconnection. The MFR then losses its equilibrium and produces the second eruption about 7 hours later. This deduction is consistent with the analysis of \citet{guo13_qmap}, who found that the QSL reconnection in the interface between the central flux and the surrounding fields, manifesting as a series of confined flares before the eruptive one, has an important role in injecting magnetic helicity and twist to the MFR. With the eruption of the MFR, the twist number and magnetic helicity in the residual flux then quickly decrease \citep{yangkai16,liur16_apj,liujiajia16}.

\subsubsection{MFR Formation during the Eruption}
It has also been proposed that the MFR can be quickly built up during the eruption via the flare reconnection. In the tether-cutting model proposed by \citet{moore01}, two oppositely sheared arcades reconnect to form a twisted loop during the onset and early phase of the eruption. In the breakout model developed by \citet{antiochos99}, the initial configuration is comprised of central sheared arcades and two neighboring flux systems. With the central flux taking off, a CS is formed below and the reconnection therein quickly transforms the sheared fluxes to the central flux to form an erupting MFR. Evidence of breakout reconnection initiating a major eruption is identified by \citet{aulanier00}. Following such an idea, \citet{macneice04} performed an MHD simulation and reproduced the complete process of the MFR eruption including the initiation, formation, and acceleration, as well as the eventual relaxation of the sheared central field to a more potential state \citep[also see][]{lynch08,karpen12}. 

Correspondingly, some observational studies support that the MFR evolving subsequently to a CME is directly formed during the eruption. The unambiguous evidence is given by \citet{liur10}, who observed that in the pre-flare phase two opposite J-shaped loops reconnect to form continuous sigmoidal loops with the central part dipped down and aligned along the PIL. Simultaneously, the compact bright loops crossing the PIL are also seen. After lasting for minutes, the sigmoidal loops quickly rise up and then produce a CME and a flare. Through observing the interaction of two pre-existing loops or filaments in the initiation phase of two eruptive flares, \citet{chenhd14} and \citet{chenhd16} also noticed that some small bright loops appeared below the interaction region and some new helical lines connecting the two far ends of the pre-existing loops are formed at the same time. They pointed out that the formation process of the helical structure basically agrees with the tether-cutting scenario.

Observations revealing the formation of the MFR during the main phase of flares are very rare. \cite{song14_formaion} reported an interesting limb eruptive event, which shows that the blob-like MFR could be built up during the CME eruption phase. It is seen that the expansion of a low-lying coronal arcade stretches the overlying magnetic field, whose legs are then curved in, forming an X-point in between. Then, the reconnection near the X-point leads to the formation and eruption of the hot boob like MFR. However, it is difficult to ensure that the erupting MFR is fully from the reconnection; it is possible that a nascent MFR (e.g., MFR seed with a strong twist but a small flux) has existed before the eruption. Moreover, a good agreement is found between the reconnection flux calculated from flare ribbons and the flux of magnetic clouds computed using in situ observations at 1 AU, which is also regarded as a strong indication of MFR formation during the flare phase \citep[e.g.,][]{lin04,qiu07,huqiang14,gopal17}.

\section{Initiation and Early Dynamics of Solar Eruptions}
Once the MFR erupts outward, it quickly forms a CME and produces flare emissions simultaneously. For flare-associated CMEs, they usually experience a three-phase evolution: the slow rise phase, impulsive acceleration phase, and propagation phase of a nearly constant velocity \citep{zhang01,zhang04}. The three phases roughly correspond to the three phases of associated flares: the pre-flare phase, rise phase, and decay phase, respectively \citep{zhang01,zhang04,qiu04,temmer08,temmer10,cheng10_buildup}, implying the coupling between the CME eruptions and the energy release of the flares through a same physical mechanism, most likely the magnetic reconnection \citep{lin00,priest02,zhang06,linjun15}.
 
At present, we are only able to forecast the likelihood of the production of CMEs/flares empirically in the light of different properties of active regions including magnetic morphology, horizontal gradient of the magnetic field, current, magnetic helicity, magnetic shear, nonpotentiality, as well as Lorentz force etc. \citep[e.g.,][European FLARECAST project]{leka03a,leka03b,falconer08,bobra14,bobra16}. An accurate determination of the onset of eruptions is still difficult, which is primarily due to the following reasons: (1) theoretically, the initiation of CMEs/flares has not been understood thoroughly, (2) validating or distinguishing the exact initiation mechanism from the possible ones observationally is a matter of great difficulty. Recently, SDO observations provide unprecedented high cadence, high resolution, and multi-wavelength data, which allow us to study the initiation of CMEs/flares in detail. Moreover, the new observations also open a window to understand the detailed formation process of CMEs, in particular those MFR-driven CMEs. In the following, we first introduce various theoretical models that are frequently used for interpreting the onset of CMEs/flares. Then, we present some observational efforts toward answering the above questions, in particular some new knowledge achieved on the early dynamics of MFR-driven CMEs.

\subsection{Initiation of the Pre-eruptive Configuration}
\subsubsection{Initiation by Magnetic Reconnection}
In terms of whether the reconnection is involved or not, the existing initiation models can be divided into two categories. The first category is reconnection-based models including the tether-cutting model \citep{moore01}, breakout model \citep{antiochos99,karpen12}, and flux emergence model \citep{chen00}. In the tether-cutting model, the key mechanism is the reconnection in the sigmoidal core field region, which transforms two groups of sheared arcades into twisted loops, thus providing an upward Lorentz force to initiate the eruption. As mentioned in Section 3.2.2, the tether-cutting reconnection in the pre-flare phase has been observed in some events \citep[e.g.,][]{liur10,chenhd14,chenhd16}.

The breakout model resorts to the reconnection taking place at the null point that exists between the central sheared flux and overlying field. The most important feature is that the reconnection site is located above the core field, rather than in the core field as stated in the tether-cutting model. The reconnection at the null point is able to remove the constraint of the overlying flux, thereby reducing the downward tension force and allowing the central flux to escape away. Theoretically, a quadrupolar structure, which includes a central sheared arcade and two neighboring loop systems with an X-point located in between is a promising structure for breakout-type eruption. Observationally, a brightening at the X-shaped structure, and some remote brightenings at the footpoints of the two neighboring fluxes, as well as the sideways motion of the lateral loops, have been seen to support the occurrence of the breakout reconnection \citep{aulanier00,gary04,ugarte07,shenyuandeng12,cheny16,reva16}. 

In the flux emergence model of \citet{chen00}, when the emerging flux emerges within the filament channel, it can reconnect with the magnetic field below the MFR. Owing to the increase of magnetic pressure, the MFR may lose its equilibrium and then rise to form a CS below it. This is similar to the tether-cutting reconnection. One the other hand, when reconnection-favored emerging flux appears and reconnects with the outer edge of the MFR, the downward tension force is reduced, making the MFR rise up. This case is similar to the lateral breakout reconnection.

\subsubsection{MFR Initiation by MHD Instabilities}
Different from the reconnection models, the other category refers to MFR-based ideal MHD models including catastrophic loss-of-equilibrium \citep{forbes91,isenberg93,lin01phd,linjun02}, kink instability \citep{torok04}, and torus instability \citep{kliem06,olmedo10}. \cite{forbes91} and \cite{isenberg93} documented that a straight MFR can lose its equilibrium in the ideal MHD process when the photospheric sources of the constraining field approach each other. The torus instability means that the expansion of the MFR tends to develop nonlinearly if the constraining field declines with height rapidly enough. For a toroidal MFR that starts to become torus unstable, the critical decay index of the overlying field is found to be 1.5 \citep{kliem06}. \citet{olmedo10} showed that the critical value is a function of the fractional number of the partial MFR with the footpoints anchored in the photosphere, i.e., a ratio between the length of the partial MFR above the photosphere and the circumference of the MFR. Interestingly, \citet{demoulin10} and \citet{kliem14_torus} proved that the torus instability is actually an equivalent description of the catastrophic loss of equilibrium of the MFR in the MHD framework. If ignoring the minor radius of the MFR, the critical decay index is 1.5 and 1 for the circular and straight MFR, respectively. However, when the MFR is deformable and as thick as the real case, their critical indices vary but slowly, typically in the range of 1.1--1.3.

The MFR with enough twist can also become unstable, an MHD process known as the kink instability. It requires that the twist number of the MFR exceeds a threshold such as 3.5$\pi$ \citep{torok04,wangym16_twist}. When the kink instability happens, the top of the MFR should slowly ascend at first if the perturbation is upward. Then, the MFR is quickly writhed by the conversion of twist into writhe, the deformation of the MFR axis, forming an inverse $\gamma$-shaped or $\Omega$-shaped structure \citep[e.g.,][]{ji03,williams05,rust05,gilbert07,guo10_index,yanxl14,hassanin16}. At the same time, the height of the MFR increases exponentially \citep{schrijver08}, which then causes the reconnection at the cross point of two MFR legs \citep[e.g.,][]{liurui09,kliem10,tripathi13}. It is worth noticing that the rapid rotation of the MFR axis is usually regarded as a condition but not a sufficient one for judging the occurrence of kink instability \citep{lynch09}.

Recently, \citet{aulanier10} compared the distinct mechanisms through a zero-$\beta$ MHD simulation. They disclosed that, prior to the eruption, flux cancellation and tether-cutting reconnection continuously work to build up the MFR and make it ascending. When rising to the critical height at which the ideal torus instability occurs, the MFR then starts to erupt. \citet{aulanier10} thus argued that the reconnection-involved processes do not trigger the eruption but act as the key mechanisms of the MFR formation and its slow rise. The mechanism that initiates the eruption of the MFR is the ideal torus instability.

\subsubsection{Validation of Torus Instability}
%%%%%%%%%%Figure 7%%%%%%%%%%%%%
\begin{figure}
\center {
\includegraphics[width=17cm]{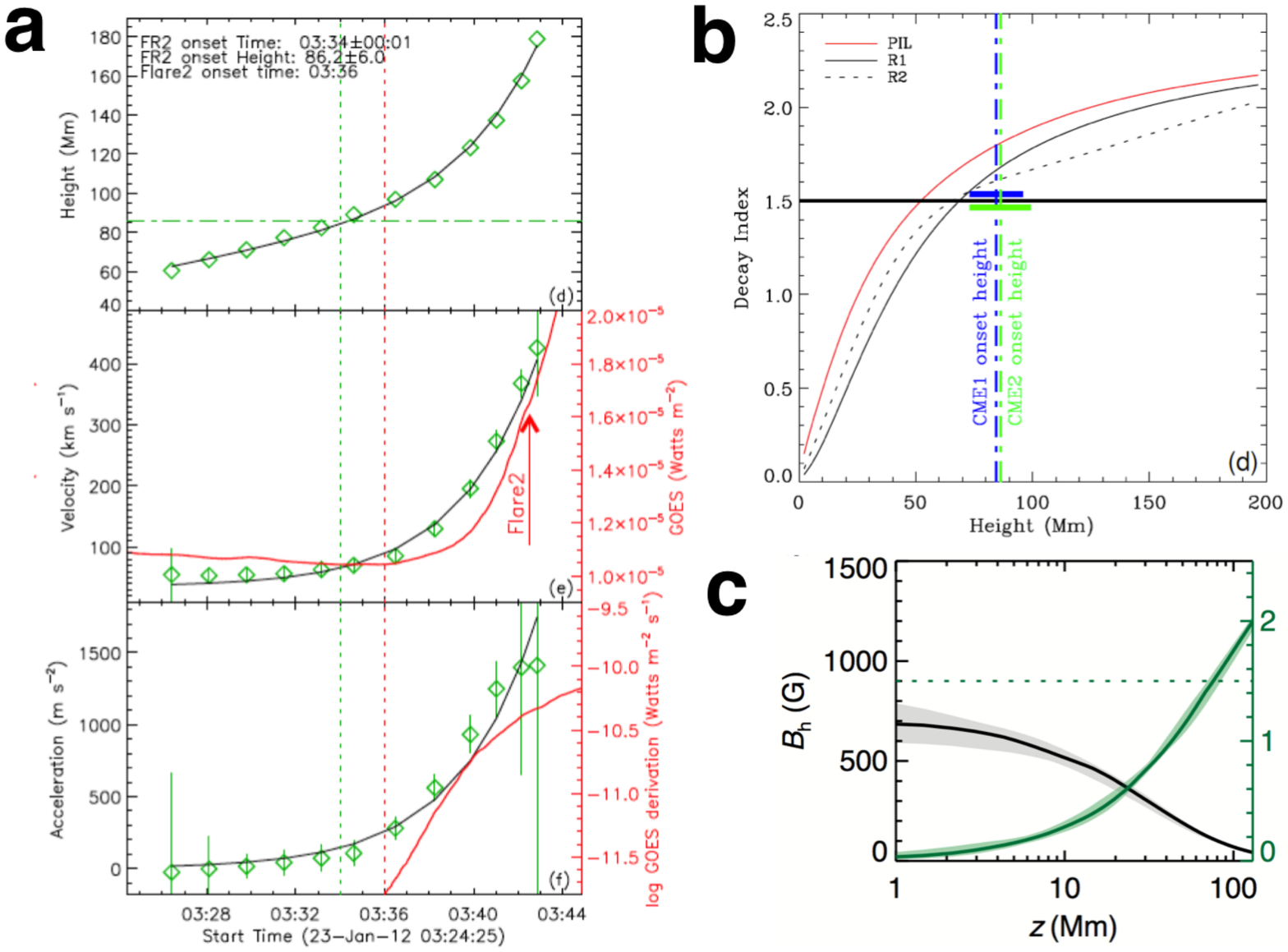}}
\vspace{-0.09\textwidth}\hspace{-0.03\textwidth}
\caption{(a) Temporal evolution of the height, velocity, and acceleration of the MFR during the early eruption with the black solid lines showing the model fitting. The red solid lines show the GOES SXR 1--8 {\AA} flux and resulting time derivation. The vertical blue (horizontal) line estimates the onset time (height) of the eruption. The vertical red line points out the onset time of the flare. (b) Distributions of the background magnetic field decay index with height over the different segments of the PIL of a CME-productive active region. The vertical lines display the onset heights of the MFR eruption with blue and green bars showing the uncertainties. The horizontal line indicates the threshold 1.5 of torus instability \citep[from][]{cheng13_double}. (c) Distributions of the background magnetic field (black) and the resulting decay index (green) with height over a CME-poor active region. The horizontal dashed line also shows the threshold 1.5 \citep[adapted from][]{sunxd15}.}
\label{onset}
\end{figure}
%%%%%%%%%%Figure 7%%%%%%%%%%%%%

In the past years, many studies attempted to validate and distinguish the right model from the available initiation models, in particular the torus instability, mainly because which can be tested from observations quantitatively. A general way is to compare the decay index of the background field at the critical height with the theoretical value. The critical height is estimated roughly as the height of the MFR just before the eruption \citep[e.g.,][]{cheng11_flare,nindos12,jiang13,chenhd14,inoue14,zuccarello14,chintzoglou15,wangrui16}. The background magnetic field is then calculated using potential field model. If the decay index of the background field at the onset height is larger than the threshold of $\sim$1.5, it is usually suggested that the torus instability plays a role in triggering the eruptions.

However, estimation of the critical height of the MFR eruption usually suffers from a significant uncertainty. In order to resolve this issue, \citet{cheng13_double} devised a mathematic model that assumes the height evolution of the MFR in the lower corona following a function $h(t)=c_{0}e^{(t-t_{0})/\tau}+c_{1}(t-t_{0})+c_{2}$, where $h(t)$ is height, $t$ is time, and $\tau, t_{0}, c_{0}, c_{1}, c_{2}$ are five free coefficients. The model consists of a linear term and an exponential term, which correspond to the slow rise phase with a constant velocity and the impulsive acceleration phase characterized by an exponential increase of velocity, respectively. Physically, this exponential term is reasonable because it describes the impulsive acceleration of the MFR \citep{schrijver08_filament} when it is triggered either by the flare reconnection \citep[e.g.,][]{antiochos99,moore01,karpen12} or by other MHD instabilities \citep[e.g.,][]{torok05,olmedo10}. Applying the mathematic model to two MFR eruption events, \citet{cheng13_double} quantitatively determined the onset time of the MFR impulsive acceleration, and found that the onset time is $\sim$2 minutes earlier than that of the associated flares (Figure \ref{onset}a). Similarly, through analysing the temporal correlation between the velocity of a filament eruption and the associated SXR emission, \citet{song15_apjl} also found that the beginning of the filament acceleration occurs earlier than that of the flare SXR emission by minutes. Combing the fact that the MFR has ascended to a height at the onset time where the decay index of the overlying field is larger than the threshold of 1.5 (Figure \ref{onset}b), it is suggested that the ideal torus instability plays a key role in initiating the impulsive acceleration of the MFR. 

Studying the magnetic environment of confined flares can also help to distinguish the distinct initiation mechanisms. A good example that has been well analysed is the flare-productive but CME-poor active region 12192, which produced 32 M-class and 6 X-class flares with only one associated with a CME. Through comparing this active region with other flare-and-CME-productive active regions, \citet{sunxd15} found that the background magnetic field in the active region 12192 is much stronger than that of others (Figure \ref{onset}c). The decay index in the lower corona (e.g., 30--100 Mm) is also smaller than the threshold of torus instability \citep[also see][]{wang07,chenhd15_12192,thalmann15,jiang16_12192}. Of course, the decay of the background magnetic field being rapid enough is not a unique condition for torus instability to take place. Another condition is the pre-existence of an MFR in the source active region \citep{liulj16}. Moreover, \citet{zuccarello17} noticed that the change of the flares from eruptive to confined is also influenced by the variation in the orientation of the pre-eruptive magnetic configuration with respect to the overlying field, rather than merely the overall change of the MHD stability.

In the past years, much attention has also been paid to the onset condition of failed eruptions. \citet{guo10_index} studied a filament eruption that firstly displays a fast rising and writhing motion but is finally confined in the lower corona. Through examining the height distribution of the decay index of the background magnetic field, they found that the decay index in the higher corona does not continuously increase, instead it starts to decrease and stays below the threshold for the torus instability, thus leading to the confinement of the filament eruption \citep[also see][]{wang07,liuyang08,cheng11_flare,joshi14,liuk15}. In addition, through a laboratory experiment, \citet{myers15} found that the confinement of the MFR eruption is also controlled by the guide magnetic field, the component of the background field that runs toroidally along the MFR axis, which interacts with electric currents in the MFR to generate a toroidal field tension force to restrict the eruption.

One should be very careful when determining the decay index at the critical height. With the help of MHD simulations, \citet{zuccarello16} found that the decay index at the height of the MFR axis is different from that at the height of the MFR top. It is suggested that the size of the MFR should not be ignored observationally when estimating the height of the MFR.

\citet{amari00,amari10,amari11} showed that flux cancellation is able to bring the initial equilibrium containing a twisted flux rope to final non-equilibrium state associated to the onset of the eruption to a critical value of free energy. They finally unified this energy criteria and the torus instability one in the case of large scale eruption \citep{amari14}.

\subsection{Early Dynamics of MFR-driven CME}

\subsubsection{Formation of the CME}
Although it is known that the eruption of various pre-eruptive structures can produce CMEs, how do they build up CMEs is still a question. The main obstacles are that (1) lack of the lower corona observations that have an enough large field of view (e.g., extending to $\sim$1.5 R$_\sun$) to guarantee the complete CME formation process observable and (2) lack of high cadence and high resolution data as the dynamical timescale of the CME formation is very short, usually of the order of minutes.   

After the launch of STEREO and SDO satellites, the above two obstacles are overcome to some extent. Using the STEREO-EUVI data, \citet{patsourakos10} studied a limb CME and found that it originates from the expansion of a plasma bubble. Shortly after the onset of the acceleration, an erupting bubble shows a fast overexpansion, which is roughly coincident with its impulsive acceleration, and it is then followed by a self-similar expansion process. The authors attributed the overexpansion to the flux conservation around a rising MFR of decreasing axial current and the flux injection to a growing MFR by the reconnection. With the high cadence SDO-AIA data, \citet{patsourakos10_genesis} found that the plasma bubble even experiences an evolution of three phases: a slow self-similar expansion, a fast but short-lived period of strong lateral overexpansion, and a self-similar expansion. They argued that it is the lateral overexpansion of the plasma bubble that creates the CME. However, they also found that the overexpansion happens during the declining phase of the flare, thus weakening the role of the flare reconnection in inducing the overexpansion. Sometimes, the overexpansion is also believed to be the origin of compression regions where type II and III bursts are produced \citep[e.g.,][]{demoulin12}. 

The discovery of the hot channel further improves our understanding of the CME formation. \citet{cheng13_driver} investigated in detail two CME events and found that the formation of the CMEs are completely controlled by the dynamics of the hot channels. In the AIA high temperature passbands, a hot channel appears as the S-shaped structure with its axis almost parallel with the PIL. After experiencing a short period of rising motion, the hot channel develops into the semi-circular structure and then quickly expands outward and speeds up. At the same time, in the AIA low temperature passbands, it is clearly seen that a plasma bubble appears and also has a fast expansion and ascending motion, very similar to the events analysed by \citet{patsourakos10} and \citet{patsourakos10_genesis}. \citet{grechnev16} also disclosed the similar formation process of a limb CME that is driven by the erupting hot MFR. Through a careful analysis, it is found that the speed of the hot channel is always faster than that of the bubble \citep{cheng13_driver}. Moreover, the hot channel not only has an overexpansion but also coincides with the overexpansion of the plasma bubble. Therefore, it is argued that early dynamics of a CME essentially depends on that of embedded hot channel, which acts as a central engine to drive the CME formation and acceleration.

\subsubsection{Emission Caused by Energetic Particles}
%%%%%%%%%%Figure 8%%%%%%%%%%%%%
\begin{figure}
\center {\hspace{0.06\textwidth}
\includegraphics[width=16cm]{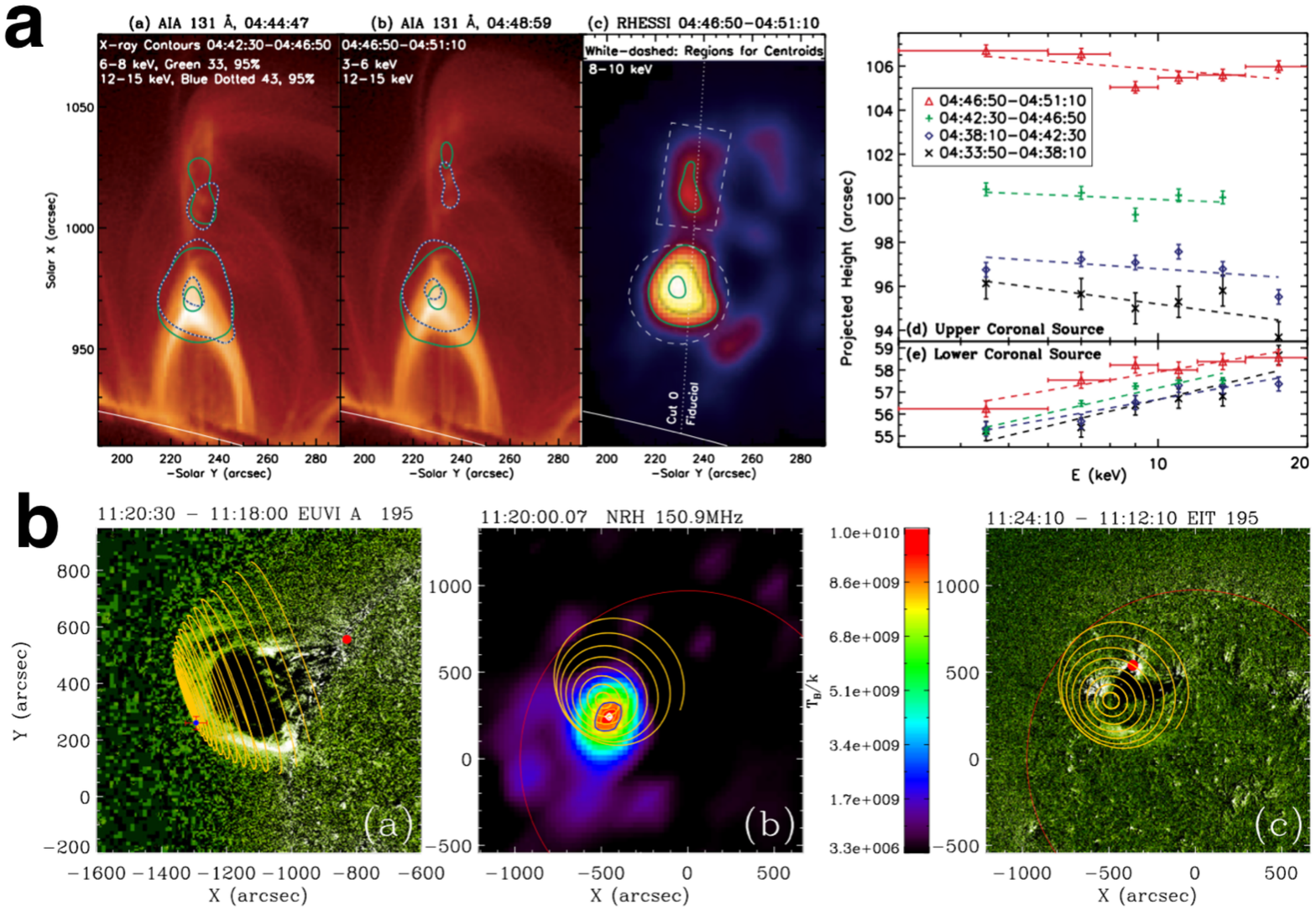}}
\vspace{-0.06\textwidth}
\caption{(a) The centroid locations and evolutions of HXR emissions induced by the erupting MFR \cite[for details please see][]{liuw13}. (b) The source of metric type II radio burst produced by the CME-driven shock \cite[for details please see][]{cheny14}.}
\label{HXR_radio}
\end{figure}
%%%%%%%%%%Figure 8%%%%%%%%%%%%%

In the acceleration phase, as expected by the standard CME/flare model \citep{carmichael64,sturrock66,hirayama74,kopp76}, the eruption stretches the overlying magnetic fields, forming a CS in the wake of the erupting MFR \citep[e.g.,][]{linjun05,linjun07,ciaravella08,cheng11_fluxrope,lileping16,zhucm16,seaton17}. The reconnection in the CS efficiently injects poloidal fluxes to the MFR and thus accelerates its eruption. The reconnection typically lasts for minutes to hours, which is mainly maintained by sink flows caused by inward-directed magnetic pressure gradient at both sides of the CS \citep{zuccarello17_vortex}. Simultaneously, the reconnection accelerates electrons, which then quickly stream down along the newly formed flare loops to heat their footpoints, mapping two parallel bright ribbons in the chromosphere \citep{forbes95,tian14_reconnection,cheng15_onset}.

To disclose the relationship between the erupting MFR and particle acceleration and further determine the location of particles acceleration, one usually needs to compare the HXR emission sources with the dynamics of CMEs/flares. \citet{liuw13} analysed the CS caused by the erupting MFR and found that both bi-directional outflows in forms of plasmoids and contracting cusp-shaped loops originate in between the hot MFR and flare loops (left panel of Figure \ref{HXR_radio}a). Moreover, these outflows are co-spatial with separated double coronal X-ray sources \citep[also see][]{sun14_dem}. The centroid separation of double coronal sources decreases with energy but increases with time (right panel of Figure \ref{HXR_radio}a). Afterwards, in the later phase of the reconnection, many dark voids are also seen to move toward the flare arcades within the CS \citep{mckenzie02,innes03,liur13_mn}. These observations show a close relationship between the erupting MFR and the production of energetic particles, suggesting that the latter may mainly occur in the reconnection outflow regions rather than in the CS.

Observations at the radio wavelength also provide an important perspective to explore where particles are accelerated during the eruption. As the MFR is accelerated continuously, a coronal shock may appear at the front of CMEs, which is proved by the appearance of metric type II radio bursts \citep[e.g.,][]{shenchenglong07,liuying09,liuying17,ma11,bain12,fengsw12,carley13}. The formation of the shock is driven by the CME expansion \citep{kouloumvakos14,cunha-silva15,wanlf16}. The average height of shocks at the onset time of type II bursts was estimated to be 0.5 solar radius \citep{gopal09_typeII} with the smallest value of 0.2 \citep{wanlf16}. The type II burst sources are usually believed to be located at the top of the shock front \citep{zimovets12,grechnev15,grechnev16}. However, through comparing the physical parameters of the shock front derived by the DEM method with that derived from the band-splitting of the type II burst, \citet{suw16} found that the sources of the type II radio burst, at least for the event they studied, are located at the flank of the shock. This result is consistent with the direct comparison of the Nancay radio images with reconstructed 3D morphology of the shock wave as done by \cite{cheny14}, who for the first time identified that the type II radio sources originate in an interaction region of the shock flank and nearby coronal ray (Figure \ref{HXR_radio}b).

Radio imaging is also a powerful tool to trace the dynamical evolution of the MFR and associated features \citep{pick05,pick08,demoulin12}. Very recently, the erupting hot MFR has been observed in the radio wavelength. With the Nobeyama Radioheliograph observation at 17 GHz, \citet{wuz16} presented the first microwave observations corresponding to a hot MFR that appears as an overall arcade-like configuration consisting of several intensity enhancements connected by weak emissions. \citet{vasanth16} even observed an obvious MFR structure in the metric wavelength and found that the associated radio emission manifests as a moving type-IV burst with their sources co-moving with the motion of the hot MFR. These observations indicate that electrons are also probably accelerated and trapped within the MFR during the eruption.

\subsubsection{3D Structure and Properties}
%%%%%%%%%%Figure 9%%%%%%%%%%%%%
\begin{figure}
\center {\vspace{-0.0\textwidth}
\includegraphics[width=17cm]{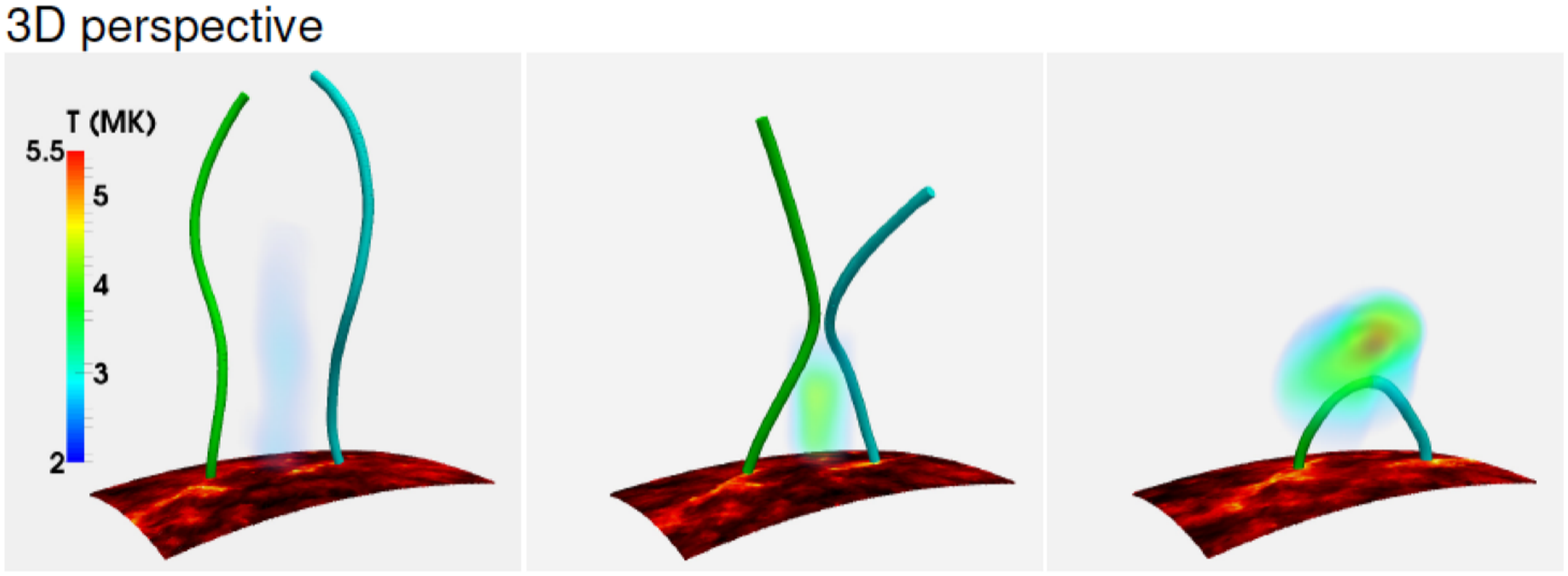}}
\vspace{-0.2\textwidth}
\caption{Reconstructed 3D topology of two magnetic field lines (cyan and green curves) before, during and after the reconnection. The bottom boundaries are projected EUVI 304 {\AA} images displaying the flare ribbons \citep[from][]{sun15_nc}.}
\label{3d_recon}
\end{figure}
%%%%%%%%%%Figure 9%%%%%%%%%%%%%

The standard CME/flare model that well interprets many aspects of the characteristics of CMEs/flares is essentially 2D. In reality, the CME/flare process is 3D \citep{aulanier12,janvier13}. \citet{cheng16_apjs} found that the axis of the pre-eruptive MFR of CMEs/flares has a significant writhing evidenced by the big ratio of its projected length to footpoint separation distance. The orientation of the axis of the MFR also significantly deviates from that of the main PIL prior to the eruption. Moreover, the flare loops are strongly sheared initially, but the shear gradually become weak with the development of CMEs/flares \citep{suyingna07a}. These characteristics show that both the pre-eruptive MFR and the background magnetic field are 3D in nature and the MFR-induced CME/flare process is also 3D. In order to understand the 3D process of CMEs/flares, \cite{aulanier12} and \cite{janvier13} extended the 2D standard CME/flare model to 3D, with which many new features can be interpreted. The strong-to-weak transition of the shear of the flare loops is found to be a result of the shear of reconnected overlying field gradually decreasing with increasing height \citep{schmieder96}.  

During the MFR eruption, the induced flare ribbons also display particular features in their morphologies and evolution. They initially appear as EUV/UV brightenings at the two ends of the MFR and then extend to two sheared J-shaped ribbons with two hooks surrounding the footpoints of the MFRs \citep{cheng16_apjs}, probably corresponding to the footprints of the curved QSLs in the chromosphere \citep{savcheva15}. At the footpoints of the MFR, both the average inclination angle and the direct current decrease with time suggestive of a straightening and untwisting of the magnetic field of the MFR legs \citep{cheng16_apjs}. These observations are basically consistent with the 3D standard CME/flare model of \cite{aulanier12} and \cite{janvier13}. Note that, however, the current at two parallel flare ribbons can be doubled as compared with the pre-flare value, contrary to that at the footpoints of the MFR, probably as a consequence of the collapse of the coronal current layer during the flare as suggested by \citet{janvier14_current} and \citet{janvier16_current}.

The morphologies of flare ribbons also depend on the three-dimensionality of the background field above the pre-eruptive MFR. \citet{sun12b} studied a non-radial MFR eruption in a quadrupolar configuration with a null point located above. Through an analysis of magnetic topology, they pointed out that the simultaneous brightening of multiple pairs of flare ribbons is a result of the reconnection between the different fluxes in the quadrupolar system \citep[also see][]{jiang14_nlfff,joshi15,zhangqm15}. In a different event, however, \citet{yang15} found that the null point can also be embedded within the quadrupolar structure. In this case, the eruption of an MFR located blow the null point leads to a three-ribbon flare with two highly elongated ones inside and outside a quasi-circular one, respectively. Very recently, a close attention is also paid to X-shaped flare ribbons. \citet{liying16} found that the flare brightenings propagate along the ribbons toward the center of an X-structure, and then spread outward in a direction more perpendicular to the ribbons. It is interpreted as the evidence of 3D reconnection that happens between two sets of non-coplanar loops that approach laterally and proceeds downward along a section of the CS. However, \citet{liur16_sr} attributed the X-shaped ribbons to the intersection of two QSL layers, i.e., the HFT, within which a separator connecting double nulls is embedded. In fact, even for most of observed flares with two parallel ribbons, the reconnection is also 3D in nature. Using the two perspectives of STEREO and SDO, \citet{sun15_nc} reported a well observed limb flare and clearly showed that two groups of field lines overlying the erupting MFR are oppositely directed and non-coplanar when they reconnect, indicating the presence of a quasi-separator. After the reconnection, the poloidal fluxes newly added to the MFR are highly helical and their two ends are still anchored in the photosphere (Figure \ref{3d_recon}).

\section{Summary and Prospects}
CMEs/flares are large-scale and most energetic eruptive phenomena in the solar system. The ejected high-speed magnetized plasma and accelerated particles may hit the Earth and thus seriously affect the safety of human high-tech activities. In the past decades, a new interdisciplinary field called space weather that refers to the solar activities, the solar wind and their influences on the magnetosphere, ionosphere, and thermosphere of our Earth has emerged. In order to understand the origin of space weather, a significant but still unsolved issue is understanding the origin and structures of CMEs/flares. In the past years, significant progresses in this aspect have been achieved. From observational perspective, we summarize the major findings and new understandings as follows:

%owing to the launch of some new satellites such as STEREO, SDO, and IRIS and the construction of some new ground-based instruments like NST, NVST, and ONSET

1. The pre-eruptive configuration of CMEs/flares is more likely to be an  MFR, which can manifest itself as a filament, filament channel, cavity, sigmoid, and hot channel etc., in dependence on the size, twist, and writhe of the MFR configuration, the viewing angle, and the wavelength at which observation is performed. In the future, we need a more advanced MHD simulation that can reproduce all these observables. In reality, the pre-eruptive configuration of CMEs/flares may not be as simple as an isolated helical MFR. Some specific characteristics such as double-decker structures and partial eruptions should also be considered.

2. The formation of the MFR is proposed either due to a direct emergence or a slow reconnection in the corona prior to the eruption, or sometimes even due to the fast reconnection during the flare. In most observations, the reconnection scenario seems a more favorable explanation for the MFR formation. MHD simulations show that the bodily emerging of a whole MFR is theoretically difficult. The reconnection is thus needed to transfer some emerged fluxes into a new MFR system in the corona prior to the eruption. However, the following questions are still unclear and need to be explored further: how does the reconnection exactly build up an MFR? what is the time scale of the MFR formation? what are the indispensable features? how much fluxes are needed to build up an unstable MFR? can we distinguish the unstable MFR from the stable one observationally?

3. The eruption of the MFR generally experiences a slow rise phase followed by an impulsive acceleration phase characterised by an exponential increase in height. The initiation mechanisms for the two phases are different and need to be clarified respectively. It is argued that the initiation of the slow rise phase could be due to diverse reasons including magnetic reconnection, MHD instabilities, and wave perturbations as long as the equilibrium of the MFR is broken. The transformation of the slow rise phase to the fast acceleration one is most likely a result of MHD instabilities. Shortly afterwards, the magnetic reconnection is then ignited to continuously accelerate the MFR eruption. However, more observations are needed to confirm this argument. 

4. The MFR is 3D in nature, e.g., a writhed hot channel-like configuration with its two elbows inclining to opposite directions and the middle part dipped toward the surface. The MFR eruption and the induced flare often show some new properties such as J-shaped and X-shaped flare ribbons, strong-to-weak change of flare loop shear, asymmetric or partial eruptions, sequential reconnection along the PIL etc. Therefore, a real MFR eruption process could be much more complex than that a 2D model predicts. In order to thoroughly understand all observables, 3D MHD simulations that consider some specific physical processes \citep[e.g.,][]{torok05,torok11,aulanier10} and 3D data-driven MHD simulations  \citep[e.g.,][]{cheung12,kliem13,fisher15,jiang16_nc}, even including the radiative transfer \citep[e.g.,][]{rempel17}, are imminently needed. Moreover, the eruption of an MFR in 3D may complicate the process of particle acceleration, which should be considered in the future as well.

\acknowledgements We are grateful to the associate editor and three anonymous referees, whose comments and suggests improved the manuscript. We also thank Prof. X.S. Feng and Prof. W.X. Wan for cordial invitation to write the review paper and the first ISSI workshop on ``Decoding the Pre­-Eruptive Magnetic Configuration of Coronal Mass Ejections" led by S. Patsourakos \& A. Vourlidas for useful discussions. X.C., Y. G. and M.D.D. are supported by NSFC under grants 11303016, 11373023,  11533005, 11203014 and NKBRSF under grant 2014CB744203.

%\bibliographystyle{apj} 
%\bibliography{reference}

\end{document}